\documentclass[prl,twocolumn,showpacs,preprintnumbers,amsmath,amssymb]{revtex4}

\usepackage{graphicx}
\usepackage{dcolumn}
\usepackage{bm}

\begin{document}

\preprint{}

\title{Solving the Cooling Flow Problem of Galaxy Clusters \\
by Dark Matter Neutralino Annihilation}

\author{Tomonori Totani}
\affiliation{%
Department of Astronomy, Kyoto University, Sakyo-ku, Kyoto, 606-8502, Japan
}%


\date{received Jan. 03, '04 and accepted Mar. 25, '04 to
Physical Review Letters}

\begin{abstract}
Recent X-ray observations revealed that strong cooling flow of intracluster
gas is not present in galaxy clusters, even though predicted
theoretically if there is no additional heating source.  I show that
relativistic particles produced by dark matter neutralino annihilation in
cluster cores provide a sufficient heating source to suppress the cooling
flow, under reasonable astrophysical circumstances including adiabatic growth
of central density profile, with appropriate particle physics
parameters for dark matter neutralinos. In contrast to other astrophysical
heat sources such as AGNs, this process is a steady and stable feedback over
cosmological time scales after turned on.
\end{abstract}

\maketitle

Diffuse thermal X-ray emission by bremsstrahlung of intracluster gas at a
temperature of $\sim$ 10 keV has been observed from galaxy clusters
for many tens of years. More
than half of clusters are called cooling flow (CF) clusters, since cooling
time of central cores is less than the Hubble time, and theorists predicted
the existence of strong CF in such systems with a rate of $\gtrsim
100 M_\odot \rm \ yr^{-1}$. However, recent X-ray observations failed to
reveal evidence of CFs, requiring that, somewhat ironically, there
must be some heat source to suppress CFs in ``cooling flow
clusters''\cite{cf-obs}. Required amount of heating is $\sim 10^{45}$ erg/s
over a time scale of the cluster age ($\sim 10^{10}$ yr).

Thermal conduction is probably playing a role, 
especially by preventing the
gas from becoming thermally unstable, but a fine tuning is necessary and
thermal conduction alone does not successfully explain all
clusters\cite{conduction-problem}. 
Another heating source popularly discussed is
active galactic nuclei (AGNs)\cite{AGN}, 
but efficiency must be very high ($\gtrsim$
10\% of the black hole rest mass energy)\cite{Fabian-L}. Generally
AGNs are intermittent activity, and
accretion rate is likely determined by dynamics of
small region around supermassive black holes 
(SMBHs). Hence it might be somewhat surprising if
all clusters are kept stable over $\gtrsim$ 100 kpc scale by feedback of
central AGNs. 

Clusters are gravitationally dominated by the cold dark matter, for which the
leading candidate is the lightest supersymmetric (SUSY) particles, plausibly
the neutralino $\chi$. 
The neutralino mass is limited in the range 30 GeV $\lesssim
m_\chi \lesssim$ 10 TeV, and thermally averaged annihilation cross section is
related to the relic density as
$\langle \sigma \upsilon \rangle \sim 3 \times 10^{-27} / (\Omega_\chi h^2)
\rm \ cm^{-3} s^{-1}$ (see, e.g., \cite{DM-SUSY-rev} for a review).
Detectability of annihilation products from high density regions such
as the Galactic center (GC)
has been widely discussed (\cite{DM-SUSY-rev,Bergstrom} and
references therein).  Here I consider a possibility that annihilation
products may contribute to heating of intracluster gas.  Exotic particle dark
matter interacting with baryons has been proposed to solve the CF
problem\cite{Qin}, but our scenario is based on theoretically
better-motivated neutralino dark matter. Neutralino annihilation in galaxy
clusters has been considered by ref.\cite{Cola}, to explain diffuse radio
halos observed in some clusters. Correlation of radio halos with merging
clusters\cite{Buote}, however, indicates that such halos are
formed by cosmic-ray electrons produced by merger shocks. 
The change
of central density profile by SMBH was not taken into account in 
ref.\cite{Cola}, 
without which the annihilation luminosity is too small to solve
the CF problem.

Since we are interested in relatively central region of a cluster, I use the
following form of the dark matter density profile at $r \lesssim r_0$: $\rho
= \rho_0 (r/r_0)^{-\gamma}$ with $r_0 =$ 0.5 Mpc and $\rho_0 =
10^{-25} \ \rm g \ cm^{-3}$, for a typical rich cluster
of $ M_{15} \equiv M_{\rm cl} / (10^{15} M_\odot) \sim 1$.  In latest
numerical simulations of cosmological structure formation, the density
profile around the center generally becomes cusps with
$\gamma \sim $ 1--1.5\cite{density}. Here I use $\gamma = 1$, 
since $\gamma = 1.5$ is not supported by a recent X-ray 
observation\cite{Lewis}.  It is easy to see that neutralino annihilation has
only negligible effects on cluster energetics, when simply this density
profile is applied.  The contribution from the cusp
to annihilation rate ($\propto \rho^2$) is 
convergent with $r \rightarrow 0$ and unimportant when $\gamma < 1.5$.
Therefore it is a good approximation to estimate the total annihilation
luminosity, $L_{\chi \chi}$, using a mean cluster density, $\rho_0$. It becomes
$L_{\chi \chi} \sim 2 \langle \sigma \upsilon \rangle \rho_0 M_{\rm cl}
c^2 / m_{\chi} \sim 2 \times 10^{40} \langle \sigma \upsilon \rangle_{-26}
m_2^{-1} M_{15} \ \rm erg/s$, where $m_2 = m_\chi / (\rm 100 GeV)$ and
$\langle \sigma \upsilon \rangle_{-26} = \langle \sigma \upsilon \rangle /
(\rm 10^{-26} cm^3s^{-1})$.  This is about five orders of magnitude lower
than that required to suppress the CF.

However, the situation drastically changes if there is a density ``spike''
associated with a central SMBH.  Adiabatic growth of a SMBH at the center of
a preexisting halo produces a spike in the density profile within $r \lesssim
r_s$ from the center, which is even steeper than original cusps\cite{GS99}.
The power-law index of spike density profile, $\gamma_s$, is related to that
of the original cusp as: $\gamma_s = (9-2\gamma)/(4-\gamma) = $ 2.33--2.4 for
$\gamma = $ 1--1.5.  The spike radius $r_s$ is a radius within which the
enclosed mass of the halo is the same as the central SMBH.  Since $\gamma_s >
1.5$, the volume integration of the annihilation rate is divergent with $r
\rightarrow 0$, and hence a huge enhancement of $L_{\chi\chi}$
is possible from the very central part of the halo.

If this enhancement is happening in the center of our Galaxy, much stronger
flux of annihilation gamma-rays and cosmic-rays are expected than previously
thought, and already existing experimental/observational limits exclude a
considerable SUSY parameter space\cite{GS99}. 
However, several authors have argued that such a process
is rather unlikely to occur in the GC\cite{Ullio-spike,Merritt}.  The
initial SMBH mass before adiabatic growth should be much smaller than the
final one, and it must be placed to the dynamical center of the halo, but the
dynamical friction time for that may be larger than the hubble time.  The
mass density of the GC is dominated by baryons rather than by dark
matter, and violent processes such as star formation and supernova explosions
might disrupt cold orbits of dark matter particles required for the spike
formation.  Mergers between halos containing SMBHs may lead to a flatter
density profile than original cusps.

However, the center of galaxy clusters, especially those having cooling cores
within the Hubble time, appear to be the best site for the spike growth to
happen. All CF clusters seem dynamically well evolved systems and a
single giant cD galaxy is placed on the gravitational center where X-ray
surface brightness peaks\cite{Jones-Forman}.  Recent Chandra observations
(e.g., \cite{Lewis}) have shown that the dark matter dominates the mass
density with a profile consistent with $\gamma = 1$ down to $\sim$ kpc,
and probably further down to the SMBH scale.
If strong CF with the theoretically
predicted accretion rate ($\sim 10^2 M_\odot \ \rm yr^{-1}$) occurred at
some epoch in early cluster evolution and
this accretion is used to feed the SMBH of a cD galaxy, the SMBH mass could
grow to $M_{\bullet, 10} \equiv M_{\bullet}/(\rm 10^{10} M_\odot) \sim 1$
within $\gtrsim 10^8$yr depending on the efficiency of mass accretion on to
the SMBH. This is much shorter than typical cluster age, while the orbital 
period at the spike radius $r_s = 1.5 M_{\bullet, 10}^{1/2}$ kpc is $5.7
\times 10^7 M_{\bullet, 10}^{1/4}$ yr, which is shorter than the SMBH growth
time scale and hence it satisfies a requirement for the adiabatic growth.
The final SMBH mass is much larger than initial mass that is probably similar
to typical SMBH mass in normal galaxies ($10^{6-9} M_\odot$), and hence
another condition of adiabatic growth mentioned by \cite{Ullio-spike} is
satisfied.

Thus, it seems reasonable to suppose that the adiabatic spike growth
occurred in the past,
within $\sim$ kpc of the cluster center where even the
latest X-ray satellites cannot resolve the density profile. 
Dominant energy production by annihilation
occurs in the central core, where the core density $\rho_c$ is limited
by annihilation itself over typical cluster age, $t_{\rm cl}
\equiv 10^{10} t_{10}$ yr, as: $\rho_c \langle
\sigma \upsilon \rangle t_{\rm cl} / m_\chi  \sim 1$. Equating
this core density and the spike density profile, $\rho =
\rho_0 (r/r_s)^{-\gamma_s}(r_s/r_0)^{-\gamma}$, 
I find the core radius
$r_c = 0.17 M_{\bullet, 10}^{2/7} \ m_2^{-3/7}  \langle \sigma \upsilon
\rangle_{-26}^{3/7} \ t_{10}^{3/7}$ pc. (It is much larger
than the Schwarzschild radius of the SMBH, 
$0.95 M_{\bullet, 10}$ mpc.)
The annihilation luminosity within this core radius is given by:
\begin{eqnarray}
L_{\chi \chi} &=& 2 m_\chi c^2 \langle \sigma \upsilon \rangle \left(
\frac{\rho_c}{m_\chi} \right)^2 \left( \frac{4 \pi}{3} r_c^3 \right)
\nonumber \\
&=& 1.9 \times 10^{44} M_{\bullet, 10}^{6/7} m_2^{-2/7} \langle \sigma \upsilon
\rangle_{-26}^{2/7} t_{10}^{-5/7} \ \rm erg \ s^{-1}.
\nonumber
\end{eqnarray}
This is a rate within $r_c$, and adding integration at $r > r_c$ 
increases the rate by a
factor of 2.8. The above equation suggests a modest time evolution of
$L_{\chi\chi} \propto t^{-5/7}$, and time average over the cluster age
is increased by a factor of 7/2.  Then I finally obtain
$L_{\chi\chi} \sim 10^{45}$ erg/s,
which is very close to the number required to heat the cooling cluster
cores. In contrast to other astrophysical heating sources,
this process is stable with only mild time evolution once it is turned
on, unless the spike is disrupted by violent events such as major
mergers. When such events destroy the spike, the cooling
core may also be destroyed, changing a CF cluster into the other
category of non-CF ones.

It must be examined whether this energy production is efficiently converted
to the heat of intracluster gas. The annihilation products are eventually
converted to stable particles. 
I used the DarkSUSY package\cite{DS} to calculate
the amount and energy spectrum of these annihilation yields, and find that,
with only weak dependence on SUSY parameters, about 1/4, 1/6 and 1/15 of the
total annihilation energy goes to continuum gamma-rays, $e^\pm$'s, and
$p\bar p$'s, respectively. The other energy goes to neutrinos, which are
not useful for heating. The spectral energy distribution of
particles per logarithmic interval,
$\epsilon^2 dN/d\epsilon$,
peaks at about 0.05, 0.05, and 0.1 times $m_\chi c^2$, in the same order.
Therefore most of the annihilation energy will be carried away by
particles of about 5 GeV, for example,
for $m_\chi \sim$ 100 GeV.

First I consider the fate of electrons/positrons. For simplicity, I assume
that a fraction $f_\pm \sim 1$ 
of the total annihilation energy is given to $e^\pm$
and all $e^\pm$'s have the same energy of $\epsilon_0 \equiv \epsilon_\pm /
(1 \ \rm GeV) \sim 1$.  They are produced at very dense environment, and
their relativistic motion would result in much higher pressure
than the environment. Therefore 
$e^\pm$'s would expand until their pressure becomes comparable with
the intracluster pressure. Buoyancy may result in 
intermittent formation of bubbles of relativistic
particles. The $e^\pm$ density in the bubbles can be
written by external pressure, as: $n_\pm \sim P_{\rm ext} / \epsilon_\pm \sim
\ 6.3 \times 10^{-7} P_{-9} \ \epsilon_0^{-1} \rm cm^{-3}$,
where $P_{-9} \equiv P_{\rm ext}/(10^{-9} \ \rm erg \ cm^{-3}) 
\sim 1$. This value and gas density $n_{-1} \equiv n/(0.1 \rm \ cm^{-3})
\sim 1$ are taken from observed values within $r \lesssim$
1--10 kpc\cite{Lewis}. 

Heating of intracluster gas by cosmic ray electrons produced by AGNs has been
discussed in literature\cite{Lea-Holman,Scott}, and similar treatments can be
applied to estimate the energy loss time scale of $e^\pm$'s, $\tau_\pm$.  The
upper limit on $\tau_\pm$ is given by the ordinary Coulomb collisions with
background gas: $\tau_{\pm, \rm cc} = 5.1 \times 10^{8} n_{-1}^{-1}
\epsilon_0$ yr.  This should be compared with radiative energy loss time
scale by inverse-Compton scattering (ICS) of the cosmic microwave background
(CMB) photons: $\tau_{\pm, \rm ic} = 1.2 \times 10^9 \epsilon_0^{-1}$ yr, and
synchrotron radiation: $\tau_{\pm, \rm sync} = \tau_{\pm, \rm ic} (B/B_{\rm
CMB})^{-2}$, where $B_{\rm CMB} = 3.30 \ \rm \mu G$. I also found that
stellar radiation energy density in cD galaxies is comparable to that of CMB,
estimating it by $U_{\rm st} \sim L_{\rm cD}/(4 \pi r_{\rm cD}^2 c)$, where
$L_{\rm cD}$ and $r_{\rm cD}$ are typical observed stellar luminosity and
spatial size, respectively\cite{Lewis}. Therefore ICS of stellar
photons should also have a comparable effect. Bremsstrahlung loss time scale
is $\tau_{\pm, \rm br} \sim 5.7 \times 10^8 n_{-1}^{-1}$ yr and annihilation
time scale of positrons is $\sim 1.0 \times 10^{10} n_{-1}^{-1} 
\epsilon_0$ yr. Comparing these time scales, it can be seen that
a considerable part of $e^\pm$ energy can be converted
into thermal energy. 

Furthermore, even more efficient energy loss is possible
when collective effects of plasma, such as the relativistic two-stream
instability, are important\cite{Scott}.  Injection of huge amount of
relativistic $e^\pm$'s within the small core radius $r_c$ would lead to
strong wind of these particles, making their momentum distribution strongly
anisotropic, which is necessary for the collective effect. Using formulae
given in these references, I found $\tau_{\pm, \rm tsi} = 3.1 \times 10^3
(n_\pm / n)^{-2} (n/{\rm cm^{-3}})^{-0.5} \epsilon_0^2 \ {\rm s} = 8.3 \times
10^6 P_{-9}^{-2} n_{-1}^{1.5} \epsilon_0^4$ yr. Dependence on $\epsilon_\pm$
is large, and almost all energy could be quickly converted to thermal energy
when $\epsilon_\pm \lesssim $ 1 GeV.

The annihilation energy given to $p\bar p$'s would also be partially
converted into thermal energy by at least the rate of Coulomb collisions that
is similar to $\tau_{\pm, \rm cc}$. Some fraction of energy may be lost by
inelastic interaction with ambient thermal protons, with a time scale of
$\tau_{\rm pp} \sim (n \sigma_{\rm 
pp} c)^{-1} = 3.3 \times 10^8 n_{-1}^{-1}$ yr,
and secondary $e^\pm$'s produced by this interaction would be again
used for heating. It is likely that primary 
annihilation gamma-rays do not contribute much to heat the cooling gas,
because the optical depth to Compton scattering is $\sim
10^{-3} \ll 1$. 
However, if there are dense clouds having large optical depth
around the annihilation core, the Compton heating may also have considerable
effect, as considered for heating by AGNs (e.g., \cite{Ciotti}).

What is the preferred value of the neutralino mass in this context?
The peak of $\epsilon^2 dN/d\epsilon$
should be less than $\sim$ a few GeV, otherwise the energy loss time scale
to heating
($\tau_{\pm, \rm cc}$ or $\tau_{\pm, \rm tsi}$) becomes longer than
that for radiative loss, leading to inefficient heating. 
Combined with the relation $L_{\chi\chi} \propto m_\chi^{-2/7}$,
the neutralino mass should be $\lesssim$ 100 GeV for the
proposed process to efficiently work. 

Now I discuss the observability of any signature of the neutralino
annihilation. Typically about $\sim$30 continuum
gamma-rays are produced at $\epsilon_\gamma >$ 
100 MeV per annihilation\cite{DS}, and expected
gamma-ray flux becomes $F_\gamma (>{\rm 100 MeV})
\sim 7 \times 10^{-8} L_{45} \ m_2^{-1}
d_2^{-2} \ \rm cm^{-2} s^{-1}$, where $L_{45} \equiv L_{\chi \chi} /
(10^{45} \rm erg \ s^{-1})$ and $d_2 = d/(100 \rm Mpc)$ is a typical distance
to nearby rich clusters.  This is just almost the same as the EGRET
sensitivity limit, and hence the prediction is marginally consistent with no
reported gamma-ray detection from nearby galaxy
clusters\cite{Reimer}. In fact, there are
many positional coincidences between known galaxy clusters and unidentified
EGRET sources, and the detection from clusters has not yet been claimed because
of low statistical significance. The
next generation gamma-ray satellite, GLAST will very likely detect the
continuum gamma-ray flux as steady and point sources at cluster centers. 

Even if the continuum gamma-rays may be discriminated from other
astrophysical sources by spectrum, variability, and/or extension, a
conclusive evidence would come from line gamma-rays.  Line flux (in $\rm
cm^{-2} s^{-1}$) is expected to be smaller than the continuum flux by a
factor of about $15 \times 10^3$, where the former factor (30/2) comes from
the ratio of photon number produced per annihilation and the latter comes
from the branching ratio into $\chi \chi \rightarrow \gamma \gamma$ or $Z
\gamma$ modes\cite{Bergstrom}.  Following the line sensitivity estimate given
in this reference, I found that the line flux expected at $\lesssim$ 100 GeV
may produce several photons from a cluster center for five-year operation of
GLAST, compared with negligibly small background rate of $\sim 10^{-3}$
events within angular and energy resolution.  Co-added analysis of many
CF clusters would even increase the sensitivity.  Future air
Cerenkov telescopes may have even better sensitivity for the line flux, but
the threshold energy must be lower than $\sim$ 100 GeV since we expect
$m_\chi \lesssim $ 100 GeV.

Since the flux enhancement by the density spike is so drastic, the best
target to search neutralino annihilation may be nearby CF
clusters, rather than the
GC or nearby galaxies. On the other hand, $L_{\chi\chi}$
without the density spike ($\sim 10^{40}$ erg/s) is below the
sensitivity limit even for the next generation gamma-ray telescopes.

Equally intriguing is the detectability of synchrotron radiation in the radio
bands, by $e^\pm$ pairs produced by annihilation.  It is known that central
cD galaxies in CF clusters have a higher probability of becoming
radio sources than those in non-CF clusters, and there seems a
correlation between radio power and CF strength
\cite{radio}, suggesting that the annihilation $e^\pm$ may be responsible for
radio emission.  Typical radio luminosity from CF cluster cores is
$\sim 10^{41-42}$ erg/s\cite{radio}, which is about $10^3$ times lower than
the cluster X-ray luminosity or $L_{\chi \chi}$.  This factor can be
explained by a few effects as follows. The magnetic field strength is
poorly known, which may be weakened by relativistic bubble formation.
Hard X-ray tails observed in some cluster spectra
are popularly interpreted as
ICS of CMB photons by cosmic-ray electrons, and their luminosity
is a few hundreds times larger than diffuse radio
halo luminosity, indicating that synchrotron is inefficient process
compared with ICS\cite{ICS-vs-sync}. The energy loss by heating can be 
much more efficient than radiative loss by the collective effect,
and then
the total radio flux will be further 
reduced. Characteristic synchrotron frequency is
$\nu_{\rm sync} \sim 0.011 \epsilon_0^2 (B/\mu\rm G)$ GHz.  Since the
injection of relativistic $e^\pm$ occurs into relatively narrow energy range
around $\sim 0.05 m_\chi c^2$ compared with broad power-law spectrum
of cosmic-ray electrons, most of radio emission may occur at $\lesssim$ GHz,
which is out of typical observing frequencies.  

Here I give a size estimate of the radio emitting region.  It is expected
that the size of $e^\pm$ bubble, $r_b$, is determined by the pressure balance
and the lifetime of these particles, as: $r_b = (3 \tau_\pm \dot N_\pm / 4
\pi n_\pm)^{1/3}$, where $\dot N_\pm = f_\pm L_{\chi \chi}/\epsilon_\pm$ is
the $e^\pm$ production rate.  Using $\tau_{\pm, \rm tsi}$ for the lifetime of
$e^\pm$'s, I found $r_b = 13 \ n_{-1}^{1/2} \epsilon_0^{4/3} P_{-9}^{-1}
f_\pm^{1/3} L_{45}^{1/3}$ kpc.  Typical extension of the radio emission
associated with cD galaxies is $\sim$ 5--10 kpc\cite{radio}. 
Large scale radio halos
($\gtrsim$ Mpc) are found only in non-CF clusters like Coma, but CF
clusters often have less extended ``minihalos'' ($\lesssim$ 100 kpc) in which
a strong radio cD galaxy is centered \cite{radio-minihalo}.  It may be
speculated that the $e^{\pm}$ bubble corresponds to the central strong radio
source, while radio minihalos are made by leaking $e^\pm$ from the bubble,
perhaps by diffusion. The morphology of such radio galaxies and minihalos in
CF clusters is poorly collimated and more spherical, compared with clear
bipolar jet-like structures generally found in radio galaxies\cite{radio}.
Such a trend may be difficult to explain if CF is suppressed by AGN jets,
while it is naturally understood if the steady, isotropic energy production
by neutralino annihilation is responsible. If this interpretation is true, it
means that we have already observed, though indirectly, the dark matter for
many tens of years!

Annihilation $e^\pm$ pairs would also produce a similar
luminosity to synchrotron radiation in the X-ray and MeV
bands by ICS of CMB and stellar photons,
respectively, but they are difficult to detect because of the strong thermal
X-ray emission and difficulty of MeV observations.

The simple estimate of $L_{\chi\chi}$ without the density spike, $\sim
10^{40}$ erg/s, is comparable to the power of diffuse radio halos in merging
clusters and hence neutralinos may be the origin, as pointed out by
ref.\cite{Cola}. A necessary condition for this is that the synchrotron must
be the dominant energy loss process. However, as argued above (and
refs. \cite{ICS-vs-sync}), observed hard X-ray tails in cluster spectra
indicate that synchrotron is rather inefficient process,
and diffuse synchrotron halo by annihilation $e^\pm$'s is likely still under
the current detection limit.

It is expected that stars in cD galaxies are also affected by the adiabatic
growth, and it might be interesting to seek for any signature in the central
luminosity density profile of cD galaxies by high resolution optical
observations. In fact, adiabatic growth by SMBHs is one of the proposed
explanations of cusps seen in central surface brightness profile of
elliptical galaxies\cite{vandenMarel}. However, there is no reason to
believe that stellar density profile should be the same as that of dark 
matter, and large elliptical galaxies
generally have flat stellar density cores\cite{vandenMarel}. A stellar
density spike should be weak and may not be detectable, if it is
formed from such a flat density core.

\begin{acknowledgments}
This work has partially been supported by a Grant-in-Aid for the
COE program for physics at Kyoto University.
\end{acknowledgments}

\end{document}